# The KLM+KLN Auger electron spectrum of rubidium in different matrices

A.Kh. Inoyatov[a,b], A. Kovalík[a,c], L.L. Perevoshchikov[a], D.V. Filosofov[a], D. Vénos[c], B.Q. Lee[d], J. Ekman[e], A. Baimukhanova[a,f]

[a] *Laboratory of Nuclear Problems, JINR, Dubna, Moscow Region, Russian Federation*
[b] *Institute of Applied Physics, National University, Tashkent, Republic of Uzbekistan*
[c] *Nuclear Physics Institute of the ASCR, CZ-25068 Řež near Prague, Czech Republic*
[d] *Department of Nuclear Physics, RSPE, The Australian National University, Canberra, ACT 2601, Australia*
[e] *Group for Materials Science and Applied Mathematics, Malmö University, 20506, Malmö, Sweden*
[d] *Institute of Nuclear Physics of the Republic of Kazakhstan, Ibragimov St. 1, 050032, Almaty, Kazakhstan*

*Keywords:* $^{85}$Rb; $^{85}$Sr; KLM-, KLN- Auger transitions; Atomic environment; Chemical shift; Multiconfiguration Dirac-Fock calculations;

**Abstract**

The KLM+KLN Auger electron spectrum of rubidium (Z=37) emitted in the electron capture decay of radioactive $^{83}$Sr in a polycrystalline platinum matrix and also $^{85}$Sr in polycrystalline platinum and carbon matrices as well as in an evaporated layer onto a carbon backing was experimentally studied in detail for the first time using a combined electrostatic electron spectrometer. Energies, relative intensities, and natural widths of fifteen basic spectrum components were determined and compared with both theoretical predictions and experimental data for krypton (Z=36). Relative spectrum line energies obtained from the semi-empirical calculations in intermediate coupling scheme were found to agree within 3σ with the measured values while disagreement with experiment exceeding 3σ was often observed for values obtained from our multiconfiguration Dirac-Hartree-Fock calculations. The absolute energy of the dominant spectrum component given by the semi-empirical approach agrees within 1σ with the measured value. Shifts of + (0.2±0.2) and - (1.9±0.2) eV were measured for the dominant KLM spectrum components between the $^{85}$Sr sources prepared by vacuum evaporation on and implanted into the carbon foil, respectively, relative to $^{85}$Sr implanted into the platinum foil. A value of (713±2) eV was determined for the energy difference of the dominant components of the KLM+KLN Auger electron spectra of rubidium and krypton generated in the polycrystalline platinum matrix. From the detailed analysis of the measured data and available theoretical results, the general conclusion can be drawn that the proper description of the KLM+KLN Auger electron spectrum for Z around 37 should still be based on intermediate coupling of angular momenta taking into account relativistic effects.

## 1. Introduction

The KLL Auger group is the most intense and the simplest (only nine basic spectrum components) among the K Auger groups. Consequently, it has been extensively studied both theoretically and experimentally in the past. With increasing energy, intensity of other K Auger groups drastically decreases while their complexity substantially increases (also due to narrower energy intervals they occupy in comparison with the KLL group). Thus in the atomic number region Z ~ 40, the total intensity of the KLM group amounts to only about 35 % of that of the corresponding KLL Auger group (see, e.g., [1,2]) and its full structure consists of 36 close lying components according to the intermediate coupling calculations [3,4] (including twelve doublets and three quartets) of very different intensities. Moreover, many components cannot be resolved experimentally in principle due to their small energy separations in comparison with natural component widths. As a result, the KLM Auger spectrum actually consists of several overlapping line groups. Even their experimental separation needs application of very high instrumental resolution power and very thin (several monolayers) radioactive sources (if the KLM Auger spectrum is studied in the radioactive decay) to prevent line broadening due to inelastic electron scattering in the source material. The predicted structure [3,4] of the KLM Auger spectrum was "confirmed" in several measurements only in the sense that some KLM lines were observed to be broader than expected on the basis of the natural widths of the atomic shells participating in the transitions and/or slightly "deformed".

Intensity of the KLN Auger group in the Z~40 region reaches only a few percent of the corresponding KLL group (see, e.g., [2]). Nevertheless, the KLN group complicated experimental investigation of the KLM Auger spectrum due to partial overlapping of these two groups. The energy interval of the overlap depends on atomic number Z and increases with it.

The complexity of the KLM+KLN Auger spectra and limitations of available electron spectroscopic technique allowed successful experimental research of these spectra mainly in the high Z region in the past. However, radiationless deexcitation of K-shell vacancies dominates in light elements where, moreover, results of the available KLM transition intensity calculations [2,3,5-7] differ substantially from each other (see also Fig. 1). The differences are partly caused by various treatment of relativistic effects and/or coupling schemes. Thus the calculations [3,5] were performed in intermediate-coupling scheme but only without consideration of relativistic effects while the calculations in jj-coupling were evaluated in both relativistic [2,7] and non-relativistic [6] approximations. In contrast, the Auger-electron energies are satisfactorily described by, e.g., widely used semi-empirical calculations [4] based on intermediate-coupling and experimental electron subshell binding energies.

So far the KLM Auger spectra of only twelve different elements in the atomic number region 18 < Z < 45 were measured in detail, namely Z=23 [8,9,10], 24 [9,10], 25 [10,11], 26 [10,12,13], 28 [14], 29 [15], 30 [16], 31 [8], 32 [17], 33 [18], 35 [19,20], and 36 [21].

There is also a lack of experimental data on the influence of atomic environments on the KLM Auger spectra especially for medium and heavy elements (i.e. involving atomic core levels). Such data are of considerable importance as for basic research in this field as for interpretation of weak effects in extremely complex experimental Auger electron spectra. Moreover, it was found, e.g., in the experimental investigations [22,23] that energies of the KLL Auger electrons are a quite sensitive probe of changes in local atomic environment. In Ref. [22], the krypton KLL Auger spectrum generated by nuclear decay of $^{83}$Rb in two different solid hosts (a bulk of a high purity polycrystalline Pt foil and a vacuum evaporated layer on the same type of Pt foil) was studied, while the KLL Auger spectrum of rubidium following the $^{83}$Sr and $^{85}$Sr decays in three different solid hosts (bulks of a high purity polycrystalline platinum and carbon foils and a vacuum evaporated layer on the same type of carbon foil) was investigated in Ref. [23]. This type of information is desirable also in some present neutrino physics experiments. In the neutrino project KATRIN [24], for example, a long-term stability of the energy scale of an electrostatic retardation β-ray spectrometer on the ±3 ppm level (i. e., ±60 meV at 18.6 keV) for at least two months of continuous measurements is required in order to achieve the intended sensitivity of 0.2 eV in searching for the electron antineutrino mass in tritium beta spectrum. In this regard, applicability of the K conversion electron line (kinetic energy of 17.8 keV) of the 32.2 keV E3 nuclear transition in $^{83m}$Kr generated in the electron capture (EC) decay of $^{83}$Rb for monitoring of the KATRIN energy scale was extensively investigated in the works [25,26]. The electron sources prepared by ion implantation of $^{83}$Rb into metallic substrates were found to be most suitable. Optimization with respect to the substrate material and implantation conditions requires additional extensive experimental investigations of the influence of local physicochemical environment of $^{83}$Rb atoms. Information on low energy electron spectra emitted in radioactive decay under real condition can also be helpful for another neutrino project, namely "Electron Capture $^{163}$Ho experiment" (ECHo) [27,28] based on high precision and high statistics microcalorimetric measurements of the $^{163}$Ho electron capture spectrum. An improvement of the theoretical description of this spectrum, in particular by investigating its modification due to the physicochemical environment of $^{163}$Ho atoms is required to search for the electron neutrino mass in the energy range below 1 eV.

Auger-electron emitting radioisotopes have always been an interest to the nuclear medicine society [29]. Cellular dosimetry of these radioisotopes largely depends on their energy spectra. However, experimental spectra of these radioisotopes are scarce, even for the K Auger electrons, and thus most dosimetry workers rely on theoretical energy spectra based on computer models that simulate the atomic relaxation [30,31]. New measurements are needed for benchmarking of these models in order to minimize the uncertainty of dosimetry calculation.

In this paper we present results of our experimental investigation of the KLM+KLN Auger electron spectrum of rubidium generated in the electron capture decay of the $^{83}$Sr ($T_{1/2}$=32.4 h) and $^{85}$Sr ($T_{1/2}$=64.9 d) radionuclides (see Figs. 2,3) embedded into different host matrices. The $^{85}$Sr sources used were prepared by vacuum evaporation on a polycrystalline carbon backing (C$_{evap}$) as well as by ion implantation at 30 keV into both high purity polycrystalline platinum (Pt$_{impl}$) and carbon (C$_{impl}$) foils. Likewise, the $^{83}$Sr source was prepared by ion implantation at 30 keV into the same Pt foil. These investigations were carried out in the frame of the development of super-stable calibration $^{83}$Rb/$^{83m}$Kr electron sources for the neutrino experiment KATRIN. It is also useful to note that the KLM+KLN Auger spectrum of rubidium (Z=37) was experimentally studied for the first time in this work.

## 2. Experimental

### 2.1. Source preparation

#### 2.1.1. Ion implantation

The $^{83}$Sr and $^{85}$Sr radionuclides were produced by spallation of metallic yttrium by 300 MeV protons from the internal beam of the synchrocyclotron particle accelerator at the JINR, Dubna. After three days of "cooling", the irradiated target (1 g weight) was dissolved in concentrated nitric acid. Using "Sr resin" (TrisKem International), strontium was chemically separated from the target material and other elements. An additional purification was carried out on a cation-exchange chromatography column (70 mm long, 2 mm in diameter, A6 resin) also in a nitric acid medium. The strontium fraction obtained ($^{82}$Sr ($T_{1/2}$=25.3 d), $^{83}$Sr ($T_{1/2}$=32.4 h), $^{85}$Sr ($T_{1/2}$=64.9 d)) was then used for both methods of electron source preparation, namely mass separation and vacuum evaporation.

The mass separation of the strontium isotopes was performed on a Scandinavian-type mass separator at the JINR, Dubna. Simultaneously, the strontium ions were embedded at the energy of 30 keV into the platinum or carbon foils. Surfaces of the foils were cleaned by alcohol before use. The places of the foils containing the Sr isotopes with the atomic mass number A = 83 and 85 were cut out and used for the electron spectrum measurements. The typical size of the "active" spot was about 2x2 mm$^2$. The activity of $^{85}$Sr in the platinum and carbon backings upon the preparation was 950 and 380 kBq, respectively and that of $^{83}$Sr in the platinum foil about 11.7 MBq.

In order to obtain some information on the depth distribution of the implanted $^{83}$Sr and $^{85}$Sr ions, we performed a Monte-Carlo simulation of ion implantations employing the computer code *SRIM* [34]. In the simulations, real circumstances of our implantations were taken into account, namely the zero ion incident angle (relative to the source foil normal), polycrystalline structure of the platinum and carbon foils as well as an adsorbed surface contamination layer represented by an additional 3 nm thick pure carbon layer on the foil surfaces [35]. Results of the simulations are displayed in Fig. 4. It is seen from the figure that in the case of the carbon foil, the $^{85}$Sr ions were embedded deep below its surface. The average ion range reaches 21.4 nm (including the 3 nm thick contamination layer). In contrast, the average $^{83}$Sr and $^{85}$Sr ion ranges in the platinum foils were calculated to be only 9.1 and 9.0 nm. Moreover, portions of about 7 and 10 % of the incident $^{83}$Sr and $^{85}$Sr ions, respectively, were found in the surface contamination layers representing different physicochemical atomic environments than those of the corresponding bulk foil material.

Corresponding amounts of ions were experimentally proved in the contamination layers in the implantation of [83]Rb into the similar Pt foil [25,26]. After the preparation, the sources were exposed to air during transfer to the electron spectrometer and the [83]Sr and [85]Sr ions in the contamination layers were thus bound with oxygen in all possible forms (oxides, hydroxides, carbonates, hydrocarbonates, etc.) and had the oxidation number +2 (as in the case of the [85]Sr source prepared by vacuum evaporation on the carbon substrate, see below the Section 2.1.2.). These contamination layers, however, were not removed from the surface of the Pt substrates (by ion sputtering or any other means) before the electron spectrum measurements.

*2.1.2. Thermal vacuum evaporation deposition*

Several drops of the strontium fraction were transferred to a Ta evaporation boat (annealed at about 1300 °C) and dried up. To remove possible volatile organic compounds from the chemical separation procedure of strontium (see Section 2.1.1.), the Ta evaporation boat with the deposited activity was first preheated at 800 °C for about 30 s. The source backing (a mechanically cleansed 150 μm thick polycrystalline carbon foil, 12 mm in diameter) was shielded all along the procedure. The source evaporation through an 8 mm diameter circular opening in a mask (fitting tightly to the foil surface) took place at 1400 °C for several seconds. In order to improve homogeneity of the evaporated layer, the source backing with the mask rotated around their common axis at a speed of 3000 turns/min at a distance of about 8 mm from the Ta evaporation boat. No visible effects were observed on the surface of the source backing after the evaporation. Two different [85]Sr sources were prepared with activities of 2.3 and 1.1 MBq just after their preparation.

The amount of the [85]Sr atoms in each of the prepared sources was several nanograms (i.e. too small to be easily examined by, e.g., the standard XPS method). It is therefore questionable to apply general chemistry terms in such cases. Moreover, the daughter [85]Rb isotope is generated in the electron capture decay of the parent [85]Sr atoms. Thus even if the "chemical state" of the [85]Sr atoms in the sources can be somehow described, it may not be the same for the [85]Rb atoms. Generally, the "chemical state" of the [85]Rb atoms can be characterized as "impurity state" in the parent [85]Sr matrices.

The exact chemical state of [85]Sr in the deposited layers in vacuum after the preparation was unknown. Because the prepared [85]Sr sources were exposed to air during their transfer to the electron spectrometer, the [85]Sr ions were bound with oxygen in all possible forms (oxides, hydroxides, carbonates, hydrocarbonates, etc. of different proportions) due to the extreme strontium reactivity and had the oxidation numbers +2. The overwhelming majority of the parent [85]Sr atoms were most likely in the $SrCO_3$ chemical form. This statement is based on: (i) specific chemical properties of strontium, (ii) its known macro-chemistry, (iii) the inner self-consistency of the physicochemical methods used for the preparation of the sources, and (iv) the conditions of their treatment. After the [85]Sr EC decay, the daughter [85]Rb atoms were stabilized in the above [85]Sr matrices. The [85]Rb ions thus were most likely bound with oxygen atoms in anions of all possible relevant forms ($O^{2-}$, $OH^-$, $CO_3^{2-}$, $HCO_3^-$, etc.). Contrary to [85]Sr, they had the oxidation number +1. It is also supported by similar experiments from the past performed, e.g., with [99m]Tc (see, e.g., [36-38]).

*2.2. Measurements and energy calibration*

An electrostatic electron spectrometer [39] was used for the electron spectra measurements. As can be seen from Fig. 5, the spectrometer combines an integral spectrometer (a retarding sphere) with differential one (double-pass cylindrical mirror energy analyzer). Several operating modes are available. In the basic mode, the scanning retarding positive voltage being applied to the electron source (1), while the retarding sphere (2) is grounded. Passing the annular conic slit (3), the slowed-down electrons enter the double pass cylindrical energy analyzer. Their energies are analyzed by the constant negative voltage (determining the absolute instrumental resolution of the spectrometer) applied to the outer coaxial cylinder (5) of the cylindrical analyzer while the inner cylinder (4) is grounded. Four circular slits (3,6) on the inner cylinder delimit the electron beam which strikes the detector (a windowless channel electron multiplier) in the second focus (F2). The detector is protected against the direct radiation from the electron source by two lead absorbers placed in the inner cylinder. The spectra were measured in sweeps. The absolute instrumental resolution and the scanning step applied depended on the intensity of the radioactive source being measured. Examples of the measured spectra are shown in Figs. 6-8.

For calibration of the spectrometer energy scale, seventeen low energy conversion electron lines (listed in parentheses) were applied. Twelve of them are emitted in nuclear transitions in $^{169}$Tm (generated in the EC decay of $^{169}$Yb ($T_{1/2}$=32.02 d)) with energies $E_\gamma$ = 8.41017(15) [41] ($M_{1,2}$, $N_{1,3}$), 20.74370(16) [41] ($L_{1-3}$, $M_{1-3}$, $N_1$), and 63.12044(3) keV [41] (K), while another five (K, $L_{1-3}$, $M_1$) in the 14.41300(15) keV [42] nuclear transition in $^{57}$Fe (originated from the EC decay of $^{57}$Co ($T_{1/2}$=271.7 d)). Energies $E_F(i)$ ($i$ is the atomic subshell index) of these calibration lines related to the Fermi level were evaluated by means of the following equation making use of the experimental Fe and Tm electron binding energies $E_{b,F}(i)$ [43] (related to the Fermi level):

$$E_F(i) = E_\gamma - E_{b,F}(i) - E_{rec}(i) \qquad (1)$$

The recoil kinetic energy $E_{rec}(i)$ of an atom after the emission of the conversion electron was calculated to be less than 0.1 eV in all our cases. Sources of the parent $^{169}$Yb and $^{57}$Co isotopes for calibration were prepared by vacuum evaporation deposition (see above Section 2.1.2.) on polycrystalline carbon substrates. Such type of sources should guarantee [44] very similar environments for the daughter $^{169}$Tm and $^{57}$Fe atoms as those for which the Tm and Fe electron binding energies [43] were determined. Nevertheless, we verified the influence of possible differences between real and tabulated [43] electron binding energies on the energy calibration. Maximum measured chemical shifts of the valence electron binding energies of about 2 and 4 eV [45] for Tm and Fe, respectively, were taken into account. The influence of these shifts on the energy calibration was found to be well below the standard deviations of the measured absolute energies of the studied electron lines quoted in Table 1 and in the text.

### 2.3. Spectra evaluation

To decompose the measured spectra into individual components, the method described in Ref. [46] was employed. The individual Auger-electron line shape was expressed as the convolution of a Gaussian (representing the spectrometer response function for monoenergetic electrons) and an artificially created function based on a Lorentzian. The Lorentzian characterizes the natural energy distribution of the investigated electrons leaving atoms. It is, however, deformed on its low energy slope due to inelastic scattering of the electrons in the source material (surface and bulk plasmon excitations, shake-up/-off effects, lattice vibrations (phonon excitations), etc.). The inelastically scattered electrons exhibit rather complicated energy structure. It consists of a wide discrete energy-loss peak (DEL, see, e.g., Fig. 6) generated mainly by electrons that suffered from surface and bulk plasmon excitations and a very long low-energy tail (going down to the "zero" energy) created by electrons which undergo multiple inelastic scattering. (In contrast, the electrons which left the electron source without any energy loss create the so-called zero-loss (or no-loss) peak which can be described by a convolution of a Gaussian and a Lorentzian resulting in the Voigt function.) Both, the position and width of the discrete energy loss peak are given by properties of the source material. However, its intensity depends on a ratio of the energy dependent mean free path for inelastic electron scattering and the effective source thickness. A description of the energy loss spectra with sufficient accuracy is a very complicated task also due to insufficient information on the measured sources including their thickness, composition, homogeneity, structure, etc. (It should be, however, noted that some progress is being made in this field, see, e.g., Refs. [35, 47-50].) Therefore, we applied the Monte Carlo approach in the spectra evaluation. It consists in manifold fitting of the measured electron spectra with random variations of the Auger-electron line shape (the same for all fitted lines in the evaluated energy interval) in its energy loss region within the pre-set shape limits. The starting approximations were created manually on the basis of suitable components of the evaluated spectrum.

In the evaluation, the fitted parameters were the position, the height, and the width (of the no-loss peak) of each Auger-electron line, the constant background and the width of the spectrometer response function. Results of the evaluations are shown in Figs. 1 and 6, 7 (continuous lines), in Tables 1-3 and in the text. The quoted uncertainties are our estimates of standard deviations ($\sigma$).

## 3. Calculations

Wave functions describing initial and final ion states in the Auger electron transitions are calculated using the relativistic atomic structure package GRASP2K [51], relying on the multiconfiguration Dirac-Hartree-Fock (MCDHF) approach. Approximate solutions to the Dirac-Coulomb Hamiltonian are generated in the framework of MCDHF theory given by atomic state functions, $\Psi(\gamma PJM)$, which are expansions over configuration state functions (CSFs), $\phi(\gamma_i PJM)$, with total angular momentum $J$ symmetry and parity $P$:

$$\Psi(\gamma PJM) = \sum_{i=1}^{N} c_i \varphi(\gamma_i PJM) \qquad (2)$$

In the expression above, $\gamma_i$ represents the configuration, coupling and other quantum numbers necessary to uniquely describe the state, $M$ is the projection of $J$ on the $z$-axis and $c_i$ are expansion coefficients. The CSFs are constructed from the one-electron Dirac orbitals, where the radial wave functions together with the expansion coefficients are obtained in a relativistic self-consistent-field procedure [52]. In the present work, electron correlation is not taken into account and the Dirac-Fock approximation within the intermediate coupling scheme is used. The transverse photon interaction as well as the leading quantum electrodynamic (QED) corrections are accounted for in subsequent relativistic configuration interaction (RCI) calculations [53].

Once a set of ASFs are obtained, the Auger transition amplitude for the autoionization of the excited intitial state $|\Psi(\gamma_i P_i J_i M_i)\rangle$ into the final scattering state is computed using RATIP [54] according to [55]:

$$V_{J_f \kappa J_i} = \langle \Psi(\gamma_t P_t J_t M_t) \| V \| \Psi(\gamma_i P_i J_i M_i) \rangle \delta_{J_i,J_t} \delta_{M_i,M_t} \delta_{P_i,P_t} \qquad (3)$$

In the expression above, $V$ denote the sum of the Coulomb electron-electron interaction and the Breit interaction. The possible final scattering states are obtained by coupling each possible state of the final ion to the wave function of the corresponding outgoing electron with relativistic angular momentum $\kappa$. Finally, the Auger transition probability is given by

$$A_{J_f \kappa J_i} = \frac{2\pi (V_{J_f \kappa J_i})^2}{\hbar} \qquad (4)$$

The frozen core approximation is used in the present calculations, and thus a common orthonormal basis with one-electron Dirac orbitals describing the initial state is used for the computation of the Auger transition probability. The energy of the outgoing Auger electron, however, is obtained as the energy difference of the initial state and the final ion state using separate orbital basis sets and is referenced to the vacuum level. The calculations were performed for vapor system, i.e., without accounting for the solid state effects. Moreover, two different creations of initial vacancies (corresponding to the real decay of the [85]Sr atoms) were taken into account, namely the electron capture (EC) decay of [85]Sr and the internal conversion (IC) processes in the [85m]Rb daughter decays. In the calculations, the atomic configuration for the IC decay was assumed to be that of a neutral rubidium system (i.e. only one electron on the 5s shell), while for the EC decay, an extra 5s electron was included, corresponding to the $5s^2$ valence-shell configuration of a neutral Sr atom. The results obtained are presented in Tables 1 and 2 and in the text.

## 4. Results and discussion

As can be seen from Fig. 6, we were able to reasonably fit only fifteen components to the measured KLM+KLN Auger spectrum. Their identification was performed on the basis of the widely used semi-empirical Auger-electron energies [4]. In some cases, however, reliability of the results obtained was greatly reduced due to a strong correlation of the fitted parameters. This concerns, in particular, complex multicomponent groups like $KL_1M_{2,3}$ and $KL_3M_1+KL_2M_{2,3}$.

It is seen from Fig. 7 that the measured dominant $KL_{2,3}M_{2,3}$ line groups exhibit very similar structure for the $^{85}$Sr sources prepared by ion implantation in Pt and C substrates despite significantly different substrate atomic numbers Z and ion implantation profiles (see Fig. 4). On the other hand, the discrete energy loss peaks of the spectrum components are much higher for the $^{85}$Sr source prepared by vacuum evaporation deposition on the carbon foil than for the above implanted sources. This difference can only be explained by a greater effective thickness for the inelastic electron scattering in the case of the evaporated $^{85}$Sr source.

### 4.1. Transition energies

In the evaluation, the absolute energy (related to the Fermi level) of the dominant and well separated $KL_3M_2(^3P_2,^3S_1)+KL_3M_3(^3P_0)$ spectral component (line No. 9) as well as energies of other spectral components relative to this one were determined. They are given in Table 1 together with results of both the semi-empirical energy calculations [4] for rubidium (4$^{th}$ -6$^{th}$ columns) and the ab initio calculations performed in the present work for our specific case, i.e., for the KLM+KLN Auger transitions in $^{85}$Rb following the electron capture decay of $^{85}$Sr (7$^{th}$ and 8$^{th}$ columns). In the case of the semi-empirical calculations [4], the relative energies are given as for the dominant components of the fitted line groups (4$^{th}$ column) as for the fitted line group energies determined as weighted mean of the semi-empirical transition energies [4] of the corresponding line group components using the theoretical transition intensities [3] (6$^{th}$ column).

It is seen from Table 1 that the relative semi-empirical energies [4] agree with the measured values within 3σ (see 5$^{th}$ column) while disagreement with experiment exceeding 3σ is often observed for values obtained from our ab initio calculations (see the last column). Moreover, the energy interval occupied by the KLM spectrum (i.e. from the first to the last spectrum lines) calculated in the present work is wider by (8.2±1.1) eV than the experimental one (485.1 eV contrary to (476.9±1.1) eV). Most of the discrepancy between the calculated energies in this work and the semi-empirical or observed energies can most likely be attributed to electron correlation effects, which are not considered in the present case (as mentioned in Section 3). It should be, however, noted that the value of 9.5 eV obtained from our calculations for the separation of the $KL_1M_2$ and $KL_1M_3$ spectrum lines (i.e. the spectrum components Nos. 2 and 3) closely matches the experimental one (9.6±0.6) eV. Values of 8.4 eV (determined from the dominant transition energies) and 8.8 eV (determined from the weighted mean energies of these two components) were obtained from the semi-empirical calculations [4]. In the case of the separation of the $KL_3M_2(^3P_2,^3S_1)+KL_3M_3(^3P_0)$ and $KL_3M_3(^3D_2,^3D_3,^1P_1)$ components (fitted lines Nos. 9 and 10), the both semi-empirical values [4] 5.9 eV (dominant transition energies) and 6.2 eV (weighted mean energies) as well as our theoretical value of 6.7 eV matched well with the experimental value of (6.1±0.5) eV. As can be seen from Table 1, the same situation is observed for the separation of the most intense spectrum components $KL_2M_3(^1D_2,^3D_1)+KL_2M_2(^3P_1)$ and $KL_3M_2(^3P_2,^3S_1)+KL_3M_3(^3P_0)$ (fitted lines Nos. 7 and 9).

As can be seen from Table 1, the absolute energy (related to the Fermi level) of the dominant fitted component (No. 9) determined as the weighted mean of the semi-empirical energies [4] and the theoretical intensities [3] of the $KL_3M_2(^3P_2,^3S_1)+KL_3M_3(^3P_0)$ Auger transitions (the 6$^{th}$ column) is identical with the absolute energy of the $KL_3M_2(^3P_2)$ Auger transition as the rates [3] for the $KL_3M_3(^3P_0)$ and $KL_3M_2(^3S_1)$ transitions reach only 15 and < 1 per cents, respectively, of the $KL_3M_2(^3P_2)$ term. This value agrees very well (within 1σ) with the measured absolute energy (also related to The Fermi level) of the fitted component No. 9. Contrary, our calculated value for the $KL_3M_2(^3P_2)$ transition is lower by (14.8±1.3) eV though it is related to the vacuum level. If the work function of 2.6 eV [56] for the polycrystalline strontium (the probable matrix of the daughter rubidium atoms) is taken into account, then the discrepancy found is enlarged up to (17.4±1.3) eV. (There are, however, sound reasons to suppose that the work function of our spectrometer should be taken into account, i.e. that one of aluminum oxide which amounts to about 4 eV). But application of the correction [4] of the Auger transition energies for the solid-state effect using the value of 6.0 eV [4] for the solid-state correction term of strontium increases our calculated energy of the $KL_3M_2(^3P_2)$ transition to 13110.8 eV, i.e. the above discrepancy is reduced to (11.4±1.3) eV. Further increase of our calculated KLM+KLN Auger transition energies in rubidium (resulting in improvement of the agreement between our energy calculations and experimental values) is expected when experimental electron binding energies rather than the ab initio binding energies (which was adopted in our calculations) are used.

Quite different situation was discovered [23] in the case of the KLL Auger spectrum of rubidium measured with the same evaporated source on the polycrystalline carbon foil. The measured absolute energy (related to the Fermi level) of the dominant $KL_2L_3(^1D_2)$ transition was found to be higher by (6.1±1.6) eV than the semi-empirical prediction [4]. A conclusion based on various facts was made that the main cause of the higher experimental transition energy stems from the so-called "atomic structure effect" (see, e.g., [57]) which was revealed in KX-rays of holmium for the first time. In this context, higher energies of the K Auger transitions following the EC decay can be explained as a result of additional screening of the daughter nucleus by a "spectator" electron because the $10^{-16}$-$10^{-17}$ s lifetime of the 1*s* atomic hole produced in the EC decay is so short that the intermediate state has an outer-electron configuration close to that of the parent atom. In X-rays, the effect is the most pronounced for rare earth elements and especially for those from the 4*f* and 5*f* groups. In the case of rubidium which belongs to the "5*s* elements", the influence of the "atomic structure effect" is expected to be less pronounced for the K Auger transition energies. According to our calculations, the absolute energies of the KLM+KLN Auger transitions following the creation of initial vacancies by the internal conversion processes in the $^{85m}$Rb daughter decays are lower by (6.9±0.1) eV than those following the electron capture decay of $^{85}$Sr. The uncertainty in the shift is the standard deviation of the differences in peak positions in folded line spectra. The folding procedure ensure that the number of dominant components is the same in the two spectra, although the number of unfolded lines differ substantially. However, the different reference level used in our calculations and the different phase of matter considered did not enable us to investigate the influence of the "atomic structure effect" on the KLM+KLN Auger spectrum emitted in the $^{83}$Sr decay.

As mentioned above (see Section 2.1), three different $^{85}$Sr sources (namely $C_{evap}$, $Pt_{impl}$, and $C_{impl}$) were prepared in order to investigate the influence of the physicochemical environment of the daughter $^{85}$Rb atoms on the KLM spectrum of Auger electrons emitted in their deexcitation. The dominant line groups of the spectra measured with these sources are compared in Fig. 7. It is seen from the figure that the positions of the two most intense components $KL_2M_3(^1D_2,^3D_1)+KL_2M_2(^3P_1)$ (No. 7) and $KL_3M_2(^3P_2,^3S_1)+KL_3M_3(^3P_0)$ (No. 9) are almost the same for the $C_{evap}$ and $Pt_{impl}$ while those for the $C_{impl}$ are slightly lower. From the energies of these two lines as well as the $KL_3M_3(^3D_2,^3D_3,^1P_1)$ one (No. 10) we determined the energy shifts between the three spectra to be ($Pt_{impl}$ - $C_{evap}$) = - (0.2±0.2) eV and ($Pt_{impl}$ – $C_{impl}$) = + (1.9±0.2) eV. These values agree within 3σ with those of - (0.7±0.1) and + (2.2±0.1) eV [23], respectively, obtained for the rubidium KLL Auger spectra following the EC decay of $^{85}$Sr in the same sources. Because of the significantly lower atomic number (and hence also much lower probability for inelastic electron backscattering resulting in reduction of the low-energy tails of electron lines), the carbon substrate would be more suitable for the super stable calibration $^{83}$Rb/$^{83m}$Kr sources in the KATRIN project than the platinum one. But the experimental data on the KLL [23] and

KLM+KLN Auger electron spectra of rubidium indicate a strong influence of at least polycrystalline carbon matrix on their absolute energies in the case of implanted sources.

*4.2. The energy difference of the $KL_3M_2(^3P_2,^3S_1)+KL_3M_3(^3P_0)$ components between Kr and Rb*

In the EC decay chain of the $^{83}$Sr isotope (see Fig. 2), the KLM+KLN Auger electrons of both rubidium and krypton are also emitted. As can be seen from the insert in Fig. 8, the corresponding Auger electron spectra were well resolved in an overview low-energy electron spectrum measured with the $^{83}$Sr source prepared by the ion implantation into the Pt foil after about four $^{83}$Sr half-lives from the source preparation. From the measured KLM+KLN spectra, a value of (713±2) eV was determined for the energy difference the $KL_3M_2(^3P_2,^3S_1)+KL_3M_3(^3P_0)$ Auger line groups between rubidium and krypton. This value agrees very well with that one of (712.8±2.0) eV determined from the absolute energies of the $KL_3M_2(^3P_2,^3S_1)+KL_3M_3(^3P_0)$ Auger line groups of rubidium (obtained in the present work with the $^{85}$Rb source, see Table 1) and krypton (measured in Ref. [21] to be (12409.1±1.5) eV using a $^{83}$Rb source prepared by the vacuum evaporation deposition on a mirror-like aluminium backing).

On the other hand, a value of 723.8 eV obtained from the semi-empirical Auger-electron energy calculations [4] is higher by (10.8±2.0) eV than the experimental difference. However, the semi-empirical absolute Auger-electron energies [4] for krypton are referenced to the vacuum level and are valid for the gas-phase system while those for rubidium were calculated for a solid and are related to the Fermi level. When a correction [4] to Auger-electron energies for a phase change was applied with the use of both the electron binding energies [22] for krypton generated in the matrix of the high purity polycrystalline platinum foil and a solid-state correction for rubidium [4], then a value of (12 413.3±2.0) eV was obtained as the semi-empirical energy of the $KL_3M_2(^3P_2,^3S_1)+KL_3M_3(^3P_0)$ Auger line group of krypton situated in a platinum matrix. This value is then higher by (4.2±2.5) eV than the above mentioned energy [21] measured for krypton created on the Al backing but still in agreement with it within 2σ. Using the modified semi-empirical value for krypton in a solid matrix, a difference of (707.3±2.0) eV is obtained for the energy difference of the $KL_3M_2(^3P_2,^3S_1)+KL_3M_3(^3P_0)$ Auger line groups of rubidium and krypton, i.e. lower by (5.7±2.8) eV than the measured one in the present work.

It should be noted that a difference of the semi-empirical $KL_2L_3(^1D_2)$ Auger-electron energies [4] between krypton and rubidium was found to be higher by (9.5±0.8) eV [22] than the experimental value for the same $^{83}$Sr source, i.e. by almost the identical value as in the present work for the $KL_3M_2(^3P_2,^3S_1)+KL_3M_3(^3P_0)$ Auger line group.

*4.3. Transition intensities*

The measured intensities of the resolved components of the rubidium KLM+KLN Auger-electron spectrum are presented in Table 2. They are related both to the total intensity of the KLM+KLN spectrum (Σ(KLM+KLN)) and the full intensity of the $\Sigma KL_{1-3}M_{1-3}$ line group ($\Sigma KL_{1-3}M_{1-3}$). In addition, experimental KLM+KLN transition intensities [21] for the nearest neighbor element Kr (Z=36) are also presented in the table since the Auger transition rates vary slowly with Z. Experimental data are compared with results of the relativistic calculations [2] in jj-coupling scheme, the non-relativistic calculations [5] in intermediate coupling (covering only the $KL_{1-3}M_{1-3}$ transitions), and the present calculations. (Results of the non-relativistic calculations [3] are not involved in the table because their results were found [5] to be less reliable than those of Ref. [5] for some transitions.)

Very good agreement is seen (within 1σ) between experimental transition rates for Rb and Kr in the case of the well separated spectrum lines including the most intense ones (fitted components Nos. 7 and 9). In the case of the close-lying unresolved spectrum lines (like the fitted components Nos. 2, 3 and Nos. 5, 6), the agreement within 1σ is observed for their summary intensities. So it can be concluded that transition intensities of the rubidium and krypton KLM+KLN Auger electron spectra fit well with each other.

A comparison between the theoretical results and the experimental data indicates that the relativistic calculations ([2] and the present work) reproduce better the measured intensities for the $KL_1M_1$ and $KL_1M_{2,3}$ lines than the non-relativistic ones. This can be attributed [58] to contributions from the relativistic effects which play an important role for the K Auger transitions resulting in two or one *s*-vacancies in any shell. On the other hand, the relativistic calculations [2] in jj-coupling fail in the intensity description of the dominant fitted components (Nos. 7 and 9). While the predictions of both the non-relativistic calculations [5] and the present calculations in intermediate coupling scheme for these two components agree with the measured data within 2σ, the relativistic values [2] jj-coupling are lower by 4σ.

It is known (see, e.g., [59]) that intensities of the $KL_1M_{2,3}$, $KL_2M_{2,3}$, and $KL_3M_{2,3}$ Auger transitions are very sensitive to the coupling of angular momenta used in the calculations. Because the calculations [2,5, this work] differ in treating of the relativistic effects, we followed a recommendation [58] and compared theoretical and experimental values for the $KL_3M_{2,3}/KL_2M_{2,3}$ transition intensity ratio in order to investigate this effect on the rubidium KLM Auger electron spectrum. It was shown in Ref. [58] that the $KL_{2,3}M_{2,3}$ transitions are negligibly influenced by the relativistic effects but the $KL_3M_{2,3}/KL_2M_{2,3}$ transition intensity ratio is very sensitive to the coupling type. As can be seen from Table 2 (the last row) and Fig. 1 (the open circle) that the $KL_3M_{2,3}/KL_2M_{2,3}$ intensity ratio derived from the intermediate coupling calculations [5, this work] agrees with the experimental value within 1σ while the jj-coupling prediction [2] differ from it by 17σ. It is, moreover, seen from Fig. 1 that new and precise experimental data on the $KL_3M_{2,3}/KL_2M_{2,3}$ intensity ratio in the atomic number region 37<Z are especially needed to find the upper Z limit where the intermediate coupling scheme should be applied in the calculations of the KLM transition rates.

The above performed comparison of the theoretical and experimental KLM transition intensities leads to the general conclusion that the MCDHF calculations accomplished in the present work are the most successful in the prediction of the KLM Auger transition rates for rubidium.

*4.4. Natural widths of the spectrum components*

Natural widths of some fitted spectrum components of the KLM+KLN Auger electron spectrum of $^{85}$Rb (measured with the $^{85}$Sr source prepared by vacuum evaporation deposition on the polycrystalline carbon foil) are compared in Table 3 with the estimated values obtained as a sum of the corresponding experimental rubidium atomic level widths based on the data [60-62]. Generally, reasonable agreement is seen between the measured and estimated values mainly due to the large uncertainties of the former ones. This finding is somewhat surprising because all fitted spectrum components are multiplets and, moreover, some of them (e.g., Nos. 4, 7, 9, 10) are expected to exhibit [3,4] very complicated structure. Thus, e.g., the dominant $KL_3M_{2,3}$ spectrum line group consists of six components [4]: the $KL_3M_2$ line is a doublet (terms $^3P_2, ^3S_1$) while the $KL_3M_3$ one is a quartet (terms $^3P_0, ^3D_2, ^3D_3$, and $^1P_1$). These six terms occupy an energy interval of 13.9 eV [4] in such a way that the $^3P_0, ^3P_2$, and $^3S_1$ terms are grouped in the 0.8 eV interval and the others in 7.9 eV with a gap of 5.2 eV between these two groups. According to the non-relativistic intermediate coupling transition intensity calculations [3], intensities of the $KL_3M_3(^3D_3, ^1P_1)$ terms are negligible (about 2% of the total $KL_3M_3$ line intensity) and those of the $KL_3M_3(^3P_0)$ and $KL_3M_3(^3P_2)$ terms amount to 39 and 59 % of the total $KL_3M_3$ transition intensity, respectively. Since the intensity of the $KL_3M_2(^3S_1)$ term is also insignificant (less than 1% [3] of the total $KL_3M_2$ transition intensity), the six $KL_3M_{2,3}$ components are reduced to three ones. As the energies of the $KL_3M_3(^3P_0)$ and $KL_3M_2(^3P_2)$ terms differ only by 0.1 eV [4] (and thus cannot be resolved experimentally in principle due to their natural widths, see Table 3), the $KL_3M_{2,3}$ line group should be seen in an experimental spectrum taken with high instrumental resolution as two lines with the following predicted "$KL_3M_3$"/"$KL_3M_2$" intensity ratios (see also Table 2, the next-to-last row): 1.09 [2] (jj-coupling), 0.20 [3] (intermediate coupling), 0.38 [this work] (intermediate coupling), 0.49 [5] (intermediate coupling) (or 0.25 after a "redistribution" of the $KL_3M_{2,3}$ terms between the fitted components Nos. 9 and 10). This conclusion is confirmed by the results of the decomposition of our spectrum (see Fig. 6) including the natural widths of the two fitted components (Nos. 9 and 10) into the $KL_3M_{2,3}$ line group which agree well with the expected values (see Table 3) for the single lines without any broadening. Taking into account this finding and reasonable (within 2σ) agreements of the "$KL_3M_3$"/"$KL_3M_2$" intensity ratios obtained from the calculations based on the intermediate-coupling scheme with the experimental value of 0.28(5) for Rb (see Table 2), one should state again that the KLM Auger electron spectrum must be described within the frame of the intermediate-coupling scheme. The same conclusion can be drawn from a detailed analysis of the $KL_2M_{2,3}$ line group.

## 5. Conclusion

An experimental investigation of the KLM+KLN Auger electron spectrum is very complicated due to its low intensity and high complexity. However, the low energy nuclear electron spectroscopy method for solid radioactive samples developed in our laboratory enabled us to perform a detailed analysis of the KLM+KLN Auger spectrum of rubidium (Z=37) following the electron capture decay of radioactive $^{83}$Sr and $^{85}$Sr incorporated in different matrices. A general conclusion was drawn from the detailed analysis of the measured data and available theoretical results that the proper description of the KLM+KLN Auger electron spectrum for Z around 37 should still be based on intermediate coupling of angular momenta taking into account relativistic effects. To find the upper Z limit for application of this approach, new precise experimental data particularly on the $KL_3M_{2,3}/KL_2M_{2,3}$ transition intensity ratio (which was found to be very sensitive to the coupling type) are needed in the atomic number region 37< Z <60. The results obtained on energy shifts of the dominant spectrum components between different matrices clearly indicate that, among others, the choice of the host matrix for super stable calibration $^{83}$Rb/$^{83m}$Kr electron sources for the KATRIN neutrino mass experiment plays an important role and should, therefore, be thoroughly investigated.

## Acknowledgement

The work was partly supported by grants GACR P 203/12/1896 and RFFI 13-02-00756.

**Table 1**

The measured energies (in eV) of the KLM+KLN Auger transitions in $^{85}$Rb (relative to the energy of the $KL_3M_2(^3P_2,^3S_1)$ +$KL_3M_3(^3P_0)$ line group) following the electron capture decay of $^{85}$Sr evaporated on a polycrystalline carbon foil ($C_{evap}$). The data are compared with results of both the semi-empirical calculations [4] and those performed in the present work (see Section 3).

| Line № | Transition(s)[a] | Experiment This work | Theory Ref. [4][b] | DL | Ref. [4][c] | This work[d] | DTW |
|---|---|---|---|---|---|---|---|
| 1 | $KL_1M_1(^1S_0,^3S_1)$ | −340.8(7)[e] | −339.0 | −1.8(7) | −339.0 | −344.7 | +3.9(7) |
| 2 | $KL_1M_2(^1P_1,^3P_0)$ | −260.5(8) | −259.4 | −1.1(8) | −259.4 | −265.5 | +5.0(8) |
| 3 | $KL_1M_3(^3P_1,^3P_2)$ | −250.9(9) | −251.0 | +0.1(9) | −250.6 | −256.0 | +5.1(9) |
| 4 | $KL_2M_1(^1P_1,^3P_0)$+ $KL_1M_{4,5}$ | −132.2(9) | −133.9 | +1.7(9) | −132.9 | −131.9 | −0.3(9) |
| 5 | $KL_3M_1(^3P_1,^3P_2)$ | −76.7(12) | −75.5 | −1.2(12) | −73.0 | −73.1 | −3.6(12) |
| 6 | $KL_2M_2(^1S_0)$ | −68.6(12) | −65.6 | −3.0(12) | −65.6 | −66.8 | −1.8(12) |
| 7 | $KL_2M_3(^1D_2,^3D_1)$+ $KL_2M_2(^3P_1)$ | −50.7(3) | −51.2 | +0.5(3) | −51.2 | −50.8 | 0.1(3) |
| 8 | $KL_1N$ | −23.1(10) | | | | −17.3 | −5.8(10) |
| 9 | $KL_3M_2(^3P_2,^3S_1)$+ $KL_3M_3(^3P_0)$ | 13121.9(13)[f] | 13120.6[f] | −1.3(13) | 13120.6[f,g] | 13107.1[h] | −14.8(13) |
| 10 | $KL_3M_3(^3D_2,^3D_3,^1P_1)$ | +6.1(5) | +5.9 | −0.2(5) | +6.2 | +6.7 | +0.6(5) |
| 11 | $KL_2M_{4,5}$ | +78.5(8) | | | | +81.9 | +3.4(8) |
| 12 | $KL_3M_{4,5}$ | +136.1(8) | | | | +140.4 | +4.3(8) |
| 13 | $KL_2N_{2,3}$ | +195.8(11) | | | | +194.3 | −3.3(11) |
| 14 | $KL_3N_1$ | +236.3(13) | | | | +236.5 | +0.2(13) |
| 15 | $KL_3N_{2,3}$ | +253.9(8) | | | | +258.7 | +4.8(8) |

[a] Lines or line groups identified by means of the semi-empirical Auger transition energies [4]. The dominant component of the multiplet is highlighted by bold.
[b] Energies of the dominant component of the individual line group related to the energy of the $KL_3M_2(^3P_2)$ component.
[c] Differences of the line group energies determined as weighted means of the semi-empirical transition energies [4] and the theoretical transition intensities [3] of the corresponding line group components.
[d] Calculations performed for the Auger transitions in $^{85}$Rb following the electron capture decay of $^{85}$Sr.

**Table 1**: (*continued*)

[e] -340.8(7) means – (340.8 ± 0.7).
[f] The absolute energy related to the Fermi level.
[g] The absolute energy of the $KL_3M_2(^3P_2,^3S_1)+KL_3M_3(^3P_0)$ line group determined as the weighted mean of the semi-empirical transition energies [4] and the theoretical transition intensities [3] of the corresponding line group components.
[h] The absolute energy related to the vacuum level.
DL - Differences between the values obtained from the semi-empirical calculations [4] for the dominant components and the experimental data.
DTW – Differences between the values obtained from the present calculations for the dominant components and the experimental data.

**Table 2**
The measured relative intensities (in %) of the KLM+KLN Auger transitions in $^{85}$Rb following the electron capture decay of $^{85}$Sr evaporated on a polycrystalline carbon foil ($C_{evap}$) compared with results of the calculations [2,5] and those performed in the present work.

| Line № | Transition(s)[a] | Experiment | | | | | Theory[c] | | |
|---|---|---|---|---|---|---|---|---|---|
| | | This work[b] | This work[c] | | Ref. [21][c] (Kr) | | Ref. [2][d] | Ref. [5] | This work |
| 1 | $KL_1M_1(^1S_0, {}^3S_1)$[e] | 8.3(5)[f] | 10.0(6) | | 9.8(3) | | 9.9 | 8.2 | 10.9 |
| 2 | $KL_1M_2(^1P_1, {}^3P_0)$ | 7.9(8) | 9.5(9) | 17.0(13) | 7.8(4) | 17.1(6) | 5.6 | 6.4 | 7.5 |
| 3 | $KL_1M_3(^3P_1, {}^3P_2)$ | 6.2(8) | 7.5(9) | | 9.3(4) | | 9.2 | 7.6 | 8.8 |
| 4 | $KL_2M_1(^1P_1, {}^3P_0)$+ $KL_1M_{4,5}$ | 3.6(5) | 4.3(6) | | 5.6(4) | | 5.7 | 4.4 | 6.3 |
| 5 | $KL_3M_1(^3P_1, {}^3P_2)$ | 4.4(8) | 5.3(9) | 9.5(13) | 5.9(2) | 8.4(3) | 8.5 | 7.7 | 7.9 |
| 6 | $KL_2M_2(^1S_0)$ | 3.5(8) | 4.2(9) | | 2.5(2) | | 1.7 | 2.4 | 2.3 |
| 7 | $KL_2M_3(^1D_2, {}^3D_1)$+ $KL_2M_2(^3P_1)$ | 20.2(11) | 24.3(15) | | 24.9(6) | | 18.2 | 25.9 | 23.6 |
| 8 | $KL_1N$ | 2.4(4) | 2.9(5) 25.8(46)[g] | | 1.4(2) 14.6(25)[g] | | 21.3[g] | | 3.0 28.7[g] |
| 9 | $KL_3M_2(^3P_2, {}^3S_1)$+ $KL_3M_3(^3P_0)$ | 23.7(15) | 28.5(19) | 36.6(24) | 28.8(12) | 35.5(13) | 20.2 | 25.2 | 24.6 |
| 10 | $KL_3M_3(^3D_2, {}^3D_3, {}^1P_1)$ | 6.7(12) | 8.1(14) | | 6.7(6) | | 22.1 | 12.3 | 9.3 |
| 11 | $KL_2M_{4,5}$ | 2.6(3) | 3.1(4) | | 3.3(2) | | 2.6 | | 3.1 |
| 12 | $KL_3M_{4,5}$ | 3.6(3) | 4.3(4) | | 4.7(3) | | 6.0 | | 4.8 |
| 13 | $KL_2N_{2,3}$ | 2.5(3) | 3.0(4) 26.9(37)[g] | | 2.1(1) 24.7(12)[g] | | 15.3[g] | | 2.6 25.3[g] |
| 14 | $KL_3N_1$ | 0.7(2) | 0.8(2) 7.5(22)[g] | | 0.6(1) 6.9(6)[g] | | 10.6[g] | | 1.0 9.2[g] |
| 15 | $KL_3N_{2,3}$ | 3.7(3) | 4.5(4) 39.8(41)[g] | | 4.1(2) 49.4(19)[g] | | 48.2[g] | | 3.8 36.8[g] |
| | | | | | | | | | |
| 10/9 | "$KL_3M_3$"/"$KL_3M_2$"[h] | 0.28(5) | | | 0.23(2) | | 1.09 | 0.49 | 0.38 |
| | $KL_3M_{2,3}/KL_2M_{2,3}$ | 1.30(5)[i] | | | 1.30(3) | | 2.13 | 1.32 | 1.31 |

**Table 2**: (*continued*)

[a] Lines or line groups identified by means of the semi-empirical Auger transition energies [4].
[b] Normalized to $\Sigma(KLM+KLN)$.
[c] Normalized to $\Sigma KL_{1-3}M_{1-3}$.
[d] For Z=36 (Kr).
[e] The dominant component of the multiplet is highlited by bold.
[f] 8.3(5) means 8.3±0.5.
[g] Normalized to $\Sigma KLN$.
[h] See Section 3.4 for explanation.
[i] The weighted mean of the values obtained also from the measurements with the $C_{impl}$ and $Pt_{impl}$ sources (see Fig. 7)

**Table 3**

Natural widths (in eV) of some fitted spectrum components of the KLM+KLN Auger electron spectrum of $^{85}$Rb emitted in the electron capture decay of $^{85}$Sr evaporated on a polycrystalline carbon foil.

| Line № | Spectrum component | This work | Estimated[a] |
|---|---|---|---|
| 1 | $KL_1M_1$ | 11.6(18)[b] | 10.7(11) |
| 2 | $KL_1M_2$ | 8.5(25) | 8.6(10) |
| 3 | $KL_1M_3$ | 7.5(25) | 8.7(10) |
| 4 | $KL_2M_1(^1P_1,{}^3P_0)+KL_1M_{4,5}$ | 9(3) | 8.3(6)[c] |
| 7 | $KL_2M_3(^1D_2,{}^3D_1)+KL_2M_2(^3P_1)$ | 5.3(7) | 6.3(5)[d] |
| 9 | $KL_3M_2(^3P_2,{}^3S_1)+KL_3M_3(^3P_0)$ | 4.9(13) | 6.1(5)[e] |
| 10 | $KL_3M_3(^3D_2,{}^3D_3,{}^1P_1)$ | 5.9(13) | 6.2(5)[f] |
| 11 | $KL_2M_{4,5}$ | 5.6(19) | 4.4(5) |
| 13 | $KL_2N_{2,3}$ | 8.5(25) | |
| 15 | $KL_3N_{2,3}$ | 3.3(11) | |

[a] A sum of natural widths of the atomic levels participating in the Auger transition.
[b] 11.6(18) means 11.6 ± 1.8.
[c] A value evaluated for the $KL_2M_1$ transition.
[d] A value evaluated for the $KL_2M_3$ transition.
[e] A value evaluated for the $KL_3M_2$ transition.
[f] A value evaluated for the $KL_3M_3$ transition.

**Figure captions**

**Fig. 1** The $KL_3M_{2,3}/KL_2M_{2,3}$ intensity ratio, as a function of the atomic number Z. Available experimental data [8-21] (see section 1) including our value for Rb (the open circle) are compared with the results of the nonrelativistic calculations [3] (NR,IC-1) and [5] (NR,IC-2) performed in the frame of the intermediate coupling scheme as well as the nonrelativistic [6] (NR,jj) and relativistic [2] (R,jj) calculations based on jj-coupling scheme.

**Fig. 2** The incomplete decay schemes of the $^{83}$Sr radionuclide [32].

**Fig. 3** The incomplete decay schemes of the $^{85}$Sr isotope [33].

**Fig. 4** The depth distributions of the $^{85}$Sr ions implanted at 30 keV into the high purity polycrystalline platinum (upper) and polycrystalline carbon (middle) foils as well as of the $^{83}$Sr ions (lower) into the high purity polycrystalline platinum foil as calculated by the computer code *SRIM* [34]. The depth values on the x-axis (represented by the foil normal) also include the thickness of the adsorbed contamination layer on the foil surfaces represented in the simulations by a 3 nm thick pure carbon layer (see Section 2.1.1.). The vertical dashed lines serve only to facilitate a comparison of the profile positions in the depth axis.

**Fig. 5** A schematic view of the combined electrostatic electron spectrometer [39]: (1) the electron source, (2) the retarding sphere, (3) the annular conic slit, (4) the inner coaxial cylindrical electrode, (5) the outer coaxial cylindrical electrode, (6) the circular slits, (F1, F2) the first and the second focuses, respectively.

**Fig. 6** An example of the KLM+KLN Auger-electron spectrum of Rb generated in the electron capture decay of $^{85}$Sr (see Fig. 3). The $^{85}$Sr source was prepared by thermal evaporation on a polycrystalline carbon foil. The spectrum was measured in fifty-four sweeps with 60 s exposition time per spectrum point in each sweep and with 7 eV instrumental resolution and 2 eV step size. Continuous lines represent results of the spectrum decomposition into components. Wide bumps observed especially on the low-energy slopes of the $KL_1M_1$ and $KL_1M_{2,3}$ Auger-electron lines (labelled as DEL) are so-called discrete energy loss peaks (see Section 2.3.).

**Fig. 7** A comparison of the rubidium $KL_{2,3}M_{2,3}$ Auger-electron line group measured with the $^{85}$Sr sources prepared by ion implantation at 30 keV into the polycrystalline Pt (upper) and C (middle) foils (cf. Section 2.1.1.) as well as by vacuum evaporation deposition (cf. Section 2.1.2.) on a polycrystalline C foil (lower). The spectra were measured with the instrumental resolution of 7 eV and the 2 eV step in different numbers of sweeps. The exposition per spectrum point in each sweep was 60 s. The dashed vertical lines serve to facilitate position comparison among the displayed spectra.

**Fig. 8** An overview low-energy electron spectrum emitted in the $^{83}$Sr decay measured with 21 eV instrumental resolution and 7 eV step and with 30 s exposition time per spectrum point after four $^{83}$Sr half-lives from the source preparation. The spectrum was not corrected for the $^{83}$Sr decay and for the spectrometer transmission drop [39,40] with increasing electron retarding voltage. In the spectrum, electrons following the EC decay of $^{83}$Sr to $^{83}$Rb and $^{83}$Rb to $^{83}$Kr are seen. In the insert, a spectrum region including a part of the KLM Auger spectrum of Kr and the full KLM Auger spectrum of Rb is shown on an enlarged scale.

**Fig. 1**

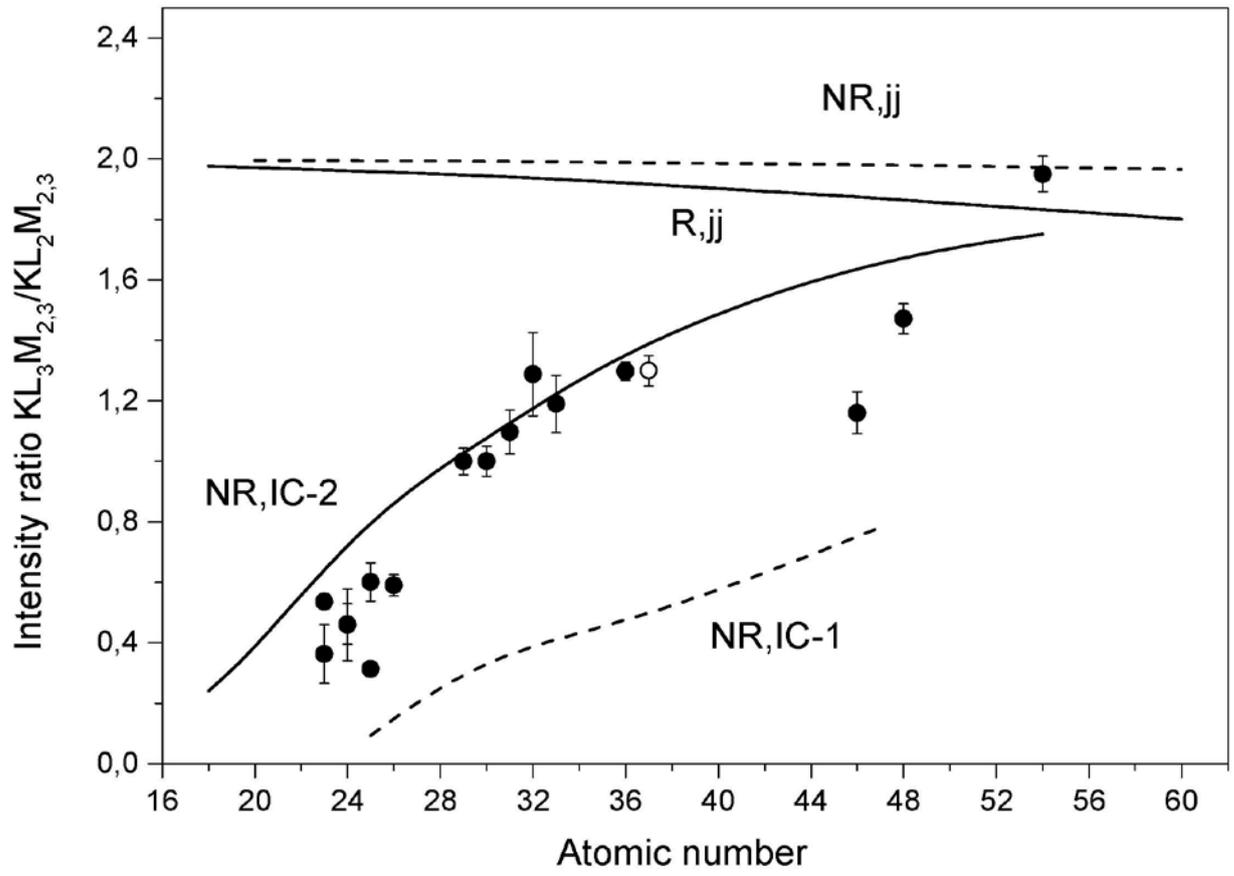

**Fig. 2**

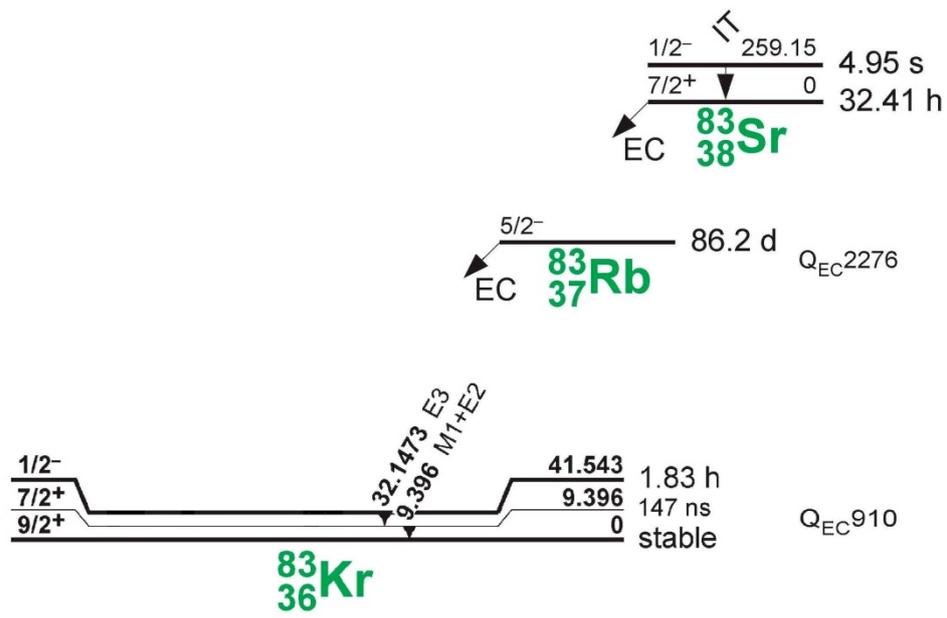

**Fig. 3**

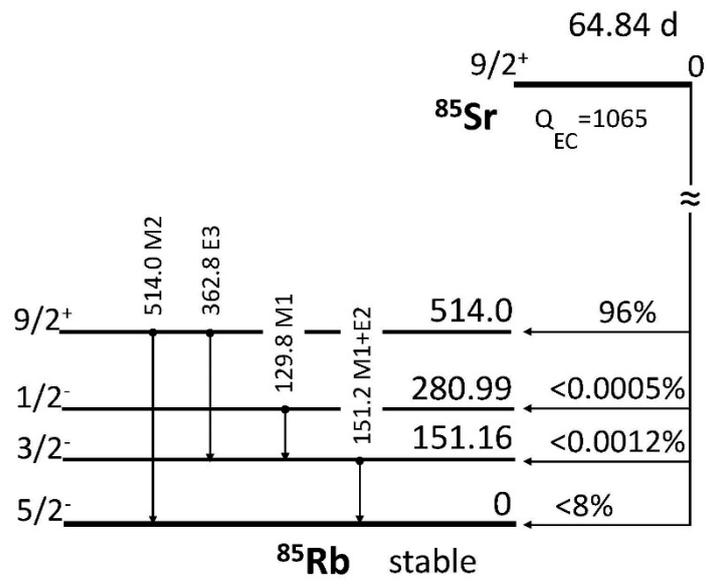

**Fig. 4**

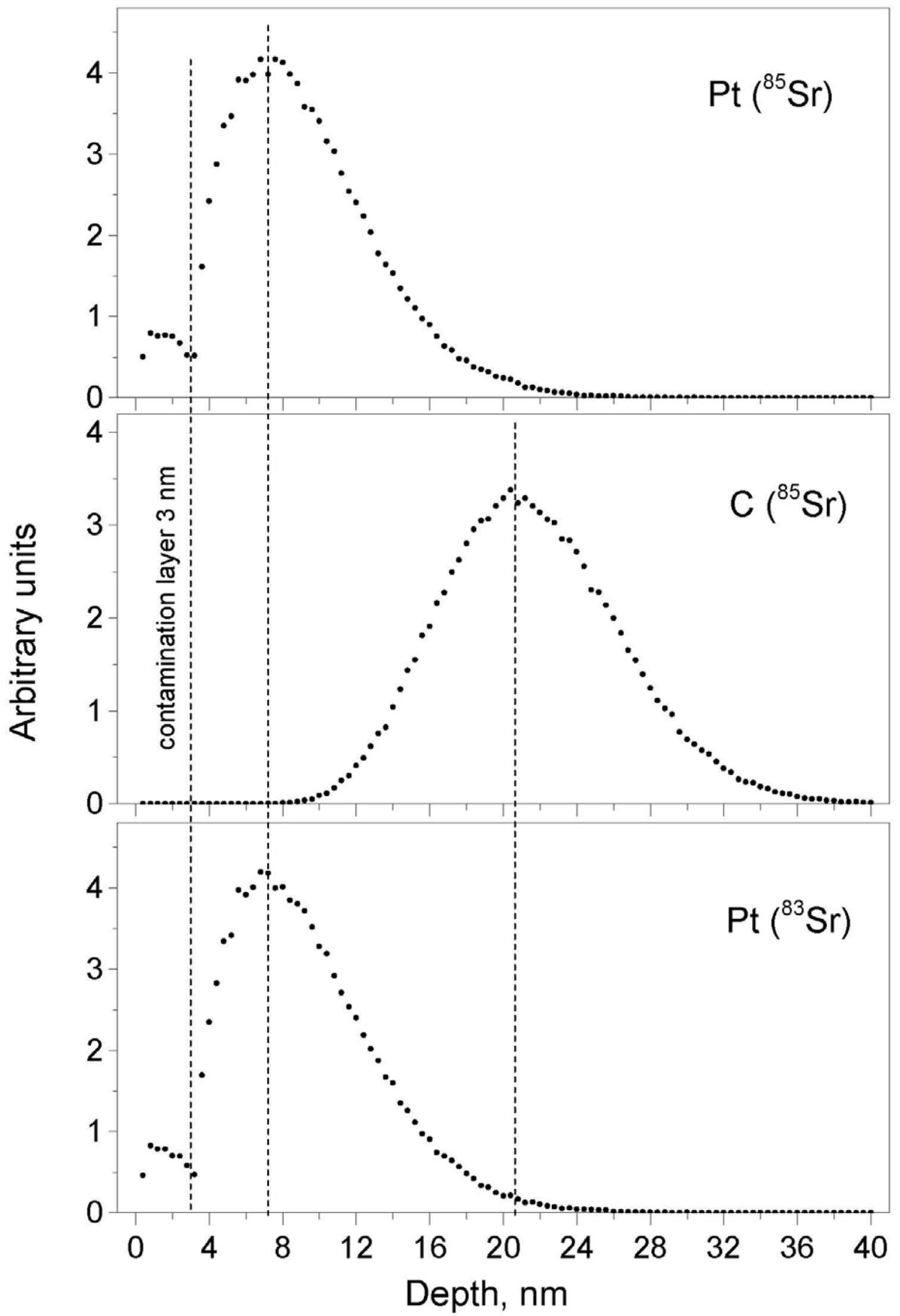

**Fig. 5**

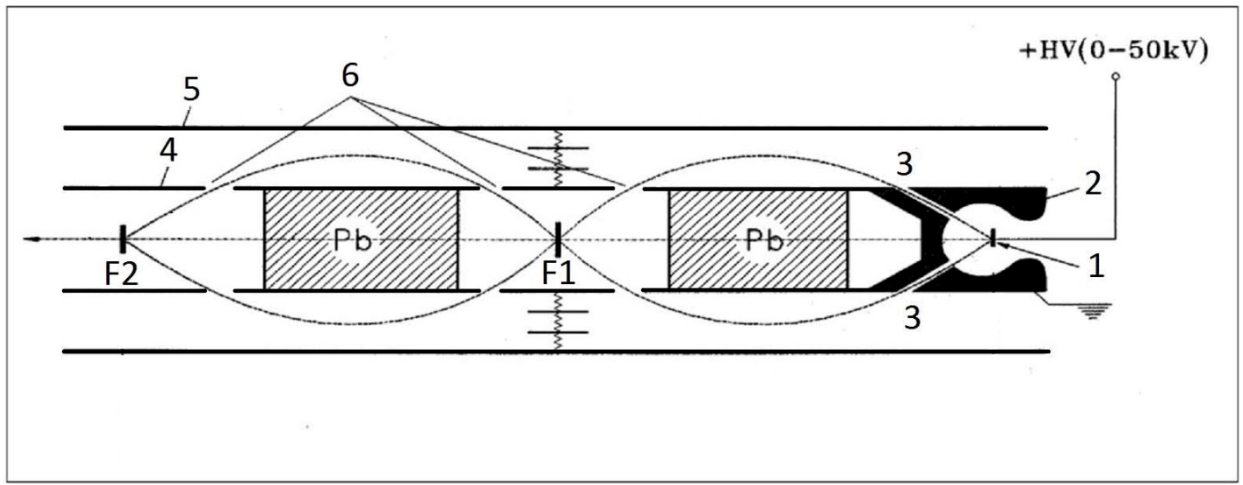

**Figure 6**

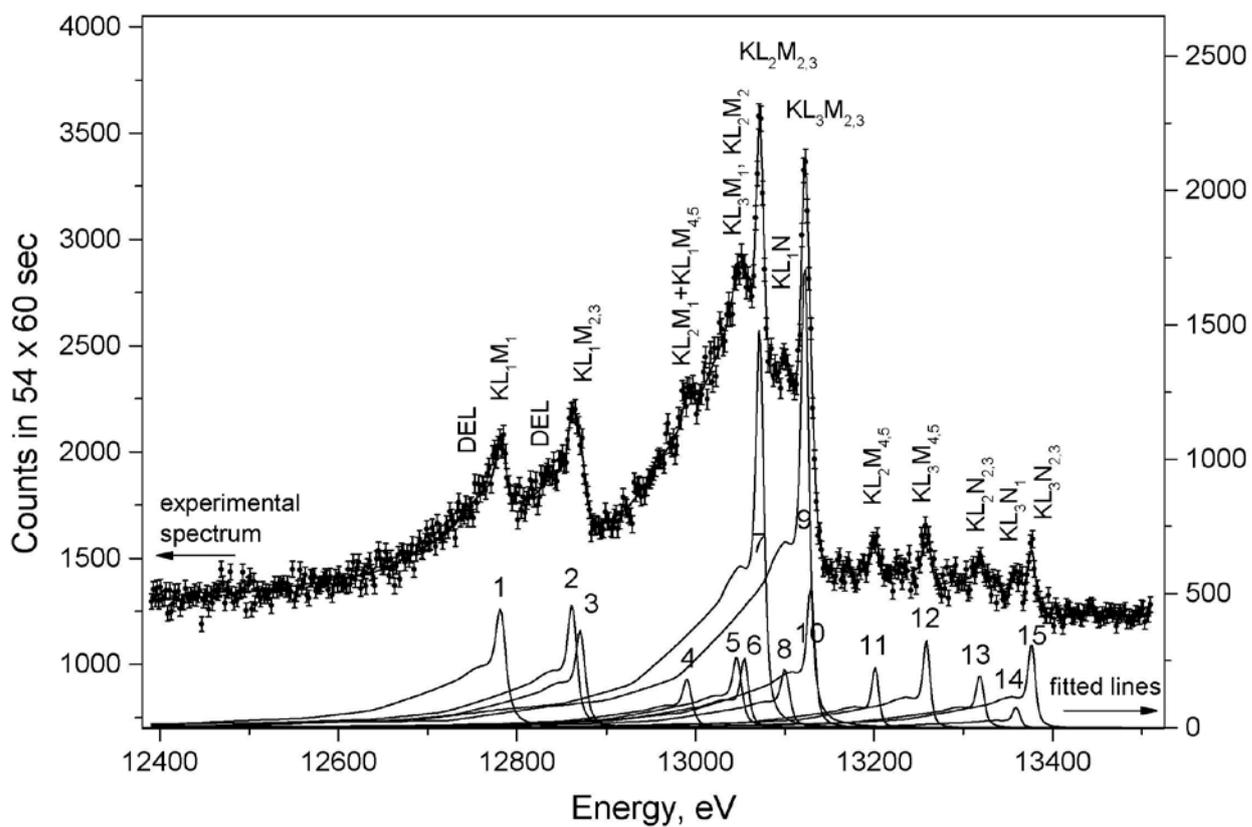

**Figure 7**

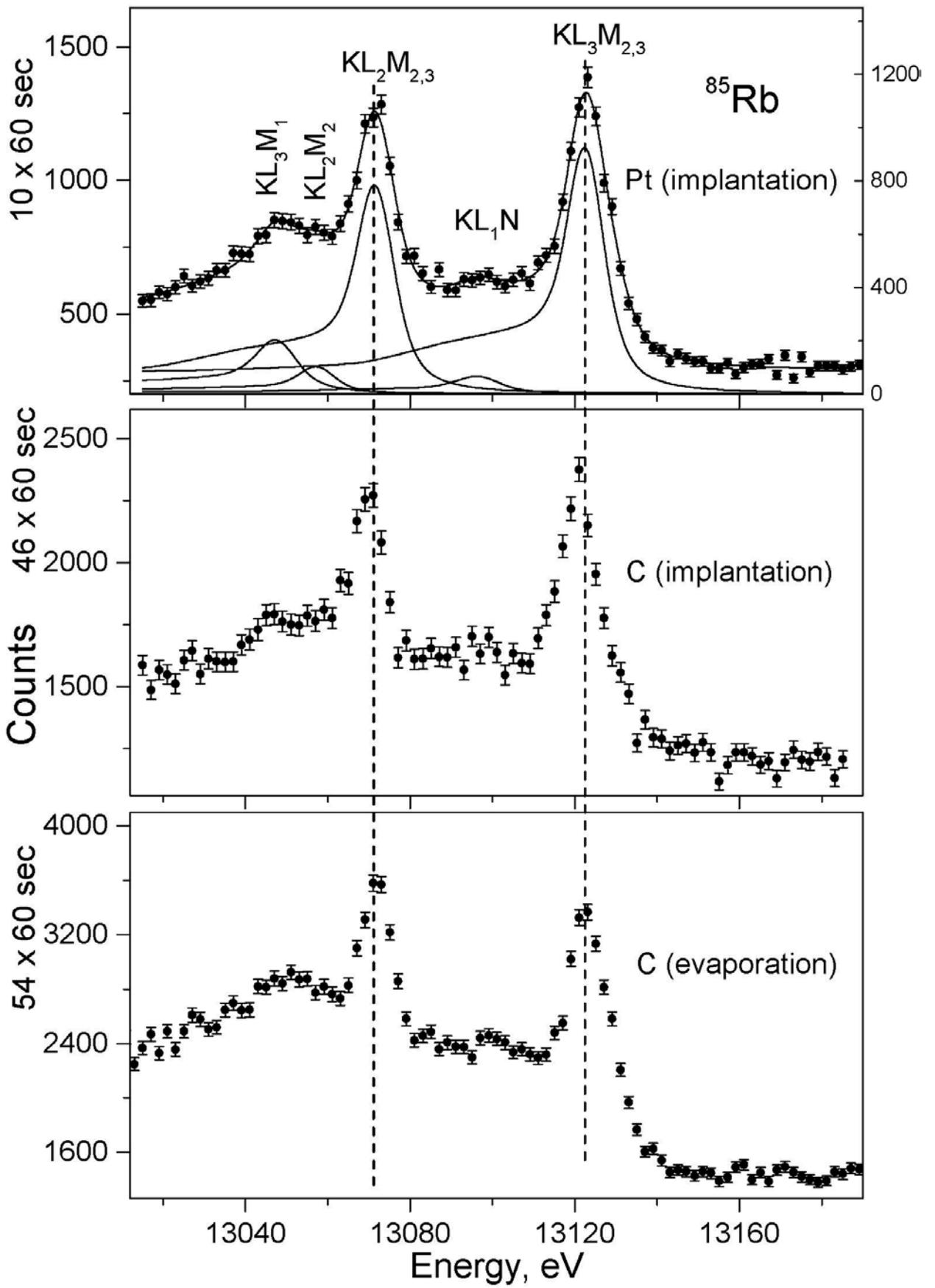

**Figure 8**